\documentstyle[aps,preprint,prl]{revtex}

\begin{document}

\title{Stripes in a three-chain Hubbard ladder: a comparison
of density-matrix renormalization group and constrained-path
Monte Carlo results} 
\author{J.\ Bon\v ca} 
\address{
Department of Physics,
FMF and J. Stefan Institute, 
Ljubljana, Slovenia}
\author {J.\ E.\ Gubernatis and M.\ Guerrero } 
\address{
Theoretical Division, 
Los Alamos National Laboratory, Los Alamos, NM 87545} 
\author{Eric Jeckelmann and Steven R. White}
\address{
Department of Physics and Astronomy,
University of California, Irvine, CA 92697}

\date{\today}

\maketitle
\begin{abstract}
Using both the density-matrix renormalization group method and the
constrained-path quantum Monte Carlo method, we have studied the
ground-state energies and the spin and hole densities of a $12
\times 3$ Hubbard model with open boundary conditions and 6
holes doped away from half-filling. Results obtained  with
these two methods agree well in the small and intermediate $U$ regimes.
For $U/t \geq 6$ we find a ground-state with stripes.
\end{abstract}


\narrowtext

One lesson learned from the search for an electronic model of high
temperature superconductivity is that computing the low lying states
of even simple many-electron models is difficult. The difficulty
in fact has prevented definitive statements about the superconducting
nature of the ground state of the one and three-band Hubbard models and the
t-J model in two dimensions.  Recently, two new methods for
computing ground-state properties of many-body systems were
developed that in a number of respects represent significant
advances over previous methods.  These are the density-matrix
renormalization group (DMRG) and the constrained-path Monte
Carlo (CPMC) methods.

The DMRG method~\cite{white92} is a variational procedure which
produces reliable estimates of the ground-state energy as well as an
exact upper bound on this energy.  It can also be used to compute
various kinds of correlation functions.  The DMRG method is extremely
accurate for low-dimensional many-body systems with short-range
interactions.  Unfortunately, it has trouble dealing with periodic
boundary conditions, and in two dimensions its computational effort
increases very quickly (probably exponentially) with the width of the
system, although only linearly with the length.  The largest DMRG
calculations for a two-dimensional doped system have been for
$16\times8$ t-J lattices\cite{white97}.  However, these calculations
were not nearly as accurate as previous DMRG calculations on narrower
ladder systems.

Like the DMRG method, the CPMC method \cite{shiwei1} produces a
variational upper bound on the ground-state energy and seemingly
accurate correlation functions.  It is a quantum Monte Carlo method
that projects the ground state from a trial state $|\Psi_T\rangle$ by
means of a branched random walk. It prevents the Fermion sign problem
from occurring by eliminating any random walker whose state
$|\phi\rangle$ develops a negative overlap $\langle
\Psi_T|\phi\rangle$ with the trial state. This constraint converts the
procedure, which is otherwise exact within statistical errors, 
into one which produces a variational upper bound on the energy.  
It can treat relatively large
systems sizes (e.g. $16\times16$ lattices), and unlike the DMRG method,
without any limitations arising from
dimensionality or boundary conditions.

While the existence of striped states was predicted nearly a
decade ago by Hartree-Fock\cite{hfdomain}, confirmation of
stripe formation with more reliable numerical techniques was
lacking. Recently, however, there has been a surge of interest 
in stripes because of their apparent experimental
observation not only in the superconducting cuprates \cite{tranquada}
but also in non-superconducting nickelates.  A stripe is simply a
domain wall ordering of holes and spins.  Spins in the regions between
the walls are anti-ferromagnetically ordered and across a wall are
correlated with a $\pi$ phase shift. The wall, at most a few lattice
spacings wide, is a hole rich region. Very recently, White and
Scalapino \cite{white97,white97c,white97b} used the DMRG method in
several studies of the existence and structure of striped states 
in the two-dimensional t-J model and in t-J ladders of several widths.
They found that stripes form under a wide variety of circumstances.
Since the t-J model is a strong-coupling approximation to
the Hubbard model, it is natural to ask if numerical methods can find
stripes in the latter model and if these states also appear in the
weak coupling limit.  It is known that within the
Hartree-Fock approximation for the Hubbard model stripes exist 
when the Coulomb interaction $U/t$ exceeds a critical value which 
is approximately 3.

In this communication we report on a numerical study of striped states
in a three-chain Hubbard model, based on both the CPMC and DMRG
methods.  The two methods complement each other, and their results
merge naturally into those of the t-J model. In this system the DMRG 
method is most effective for large $U$, while the CPMC method is most
effective for small $U$.  We find the two methods 
agreeing pleasingly well in both
the small and intermediate $U$ regimes. We do not find stripes
at small $U$; at $U/t \sim 6$ stripes appear and the Hubbard
results map nicely onto previous DMRG t-J results.

We used the Hubbard model
\begin{equation}
  H = -t\sum_{\langle i,j\rangle,\sigma}(c^\dagger_{i\sigma}c_{j\sigma}
  + c^\dagger_{j\sigma}c_{i\sigma}) + U \sum_i
  n_{i\uparrow}n_{i,\downarrow}
\end{equation}
defined on a {\it rectangular} lattice with {\it open} boundary
conditions. Open boundary conditions are used to break the
translational invariance of the system, allowing striped
patterns to appear as local density variations. 
Open boundary conditions also make the DMRG method more accurate.
We report results on a $12 \times 3$ three-chain ladder, doped six
holes away from half-filling, as a function of $U/t$.
This system was selected based on t-J results showing two
pronounced 3-hole transverse stripes\cite{white97b}.

For different values of $U$, we computed the ground-state energy
and ground state expectation values of
the rung spin density
\begin{equation}
  S(i)=\sum_j (-1)^{j-1}\langle s^z_j(i)\rangle ,
\end{equation}
and the rung hole density 
\begin{equation}
  R(i) = \sum_{j\sigma} \langle 1-n_{j\sigma}(i)\rangle,
\end{equation}
Here $i$ labels the rung, and $j$ is the chain index.
We remark that in all of our results $\sum R(i) = N_h = 6$.

The CPMC method uses a trial wave function $|\Psi_T\rangle$ for three
purposes: as a starting point, an importance function, and a
constraining function.  The first use influences the Monte Carlo
dynamics; the second, the statistical error of the results; and the third,
the systematic error.  In the simulations reported here, we used two
types of trial functions. One is the wave function of the
non-interacting ($U=0$) problem, the free-electron (FE) solution.
This single Slater determinant state does not exhibit a striped
pattern.  The second is the unrestricted Hartree-Fock (UHF) solution.
This single Slater determinant shows a striped pattern provided $U/t$ is
approximately 3 or larger. We note that for $U/t=3$ the value of the
overlap $\langle \mbox{UHF}|\mbox{FE}\rangle$ is small ($\sim 0.1$)
and becomes even smaller ($\sim 10^{-5}$) as $U/t$ is increased to $8$.

In our ``quick and dirty'' method of obtaining the UHF solution, each
$\langle n_{i\sigma} \rangle$ was obtained independently without
imposing the point group symmetry of the rectangle.  Accordingly the
UHF wave function did not precisely display the symmetries of the
rectangle, with the mirror symmetry, $y \leftrightarrow -y$,
being noticeably absent. If we used multiple Slater determinants for
$|\Psi_T\rangle$ that restored the broken mirror symmetry, either as a
two determinant trial function with the UHF solution and its mirror
symmetric solution or as a four determinant solution obtained by
applying the four group operations of the rectangle to the UHF
solution, we would find the expectation value of the energy and the
rung densities virtually unchanged. The main difference is the
symmetrized UHF state has $\langle s^z_j(i)\rangle=0$; the unsymmetrized
state does not. In the results reported below we
used the single Slater determinant UHF state as our constraining and
initial wavefunctions.
 
We remark that we used the UHF solution for $U/t=3$ even though the
simulation was for a Hubbard model with a larger $U$. The smaller $U$
starting point, a point at which a striped state in the UHF
wavefunction has become easily
noticeable, produced expectation values with less variance than
results obtained by starting with the UHF solution for the value of
$U$ in the simulation.  The smaller $U$ starting point also tended to
produce less localized stripes.  The Hartree-Fock solution for the
larger values of $U$ has a more spatially restricted domain wall,
eventually becoming one rung of lattice sites.  When we put the FE
wave function as $|\Psi_T\rangle$, the final result for any $U$ does not have
stripes, and when we put UHF wavefunctions as $|\Psi_T\rangle$, it
does for $U/t \geq 5$. 
The CPMC simulations generally used 500 random walkers (on the
average).  We estimate that the systematic error due to the
Trotter approximation was within the statistical error.

We also remark that finding a striped ground state is not simply a matter
of starting the calculations with a striped state. The situation is
more subtle. For example, in the CPMC simulations the same
wavefunction does not have to be used as the initial and the
constraining states. If the constraining state is chosen to be a
linear combination of the FE and UHF wavefunctions, then within
statistical error the same ground-state energy and non-striped hole
and spin density are found irrespective of whether the FE or UHF
wavefunction was used as the initial state. 

Unlike the Monte Carlo method the DMRG method does not have a
statistical error which is reduced by increasing the length of
simulations.  Instead it has a truncation error which is reduced by
keeping larger numbers $m$ of density matrix eigenstates (for more
details see \cite{white92}).  Varying $m$ allows one to compute
physical quantities for different truncation errors and thus to obtain
error estimates on these quantities.  In Fig.~\ref{fig1} we show
the DMRG ground-state energy as a function of the truncation error
parameterized by $m$.  For large $m$ the energy $E(P_m)$ decreases
almost linearly with the truncation error $P_m$.  Thus, we can
extrapolate to the limit $P_m \rightarrow 0$ to estimate the
ground-state energy $E(P_m=0)$ more accurately.  Usually the slight
deviation from linearity in the function $E(P_m)$ has a positive
second derivative, so extrapolated values usually increase slightly
when the accuracy is increased.  Thus our results for
$E(P_m\rightarrow 0)$ are likely to be slightly lower than the exact
results.  Based on observation of the dependence of the rung densities
on $m$, we estimate that the errors in the rung densities are
generally no more than 1\%.

As for the CPMC method, the starting wavefunction used by the DMRG
method can be varied.  This starting wavefunction is simply the DMRG
approximation to the ground state at the end of the warmup sweep,
before the finite system sweeps begin.  After these initial sweeps the
initial wavefunction has no effect.  However, it may be that there are
a few ``metastable'' ground states, in which case, with a poor initial
wavefunction, the DMRG method can become stuck for a significant
number of sweeps. If $m$ is increased enough, the method will always
find the correct ground state, but the calculation may not
always be practical.  Thus it is often wise, particularly in 2D
systems, to try several different initial wavefunctions
\cite{white97}. Usually we generate distinct initial
wavefunctions by varying the quantum numbers of the target state
as the system is built up from scratch.  This crude approach is
usually sufficient to avoid getting stuck in metastable states.
For the system studied here, it was feasible to increase $m$
enough to tunnel out of metastable ground states, which allowed
a check on whether we had tried enough initial wavefunctions.

Our main results for the energies are reported in Table~1.  Two sets
of CPMC results are given. The labels FE and UHF indicate the choice
of $|\Psi_T\rangle$.  For $U/t$ less than about 8, the FE CPMC energies lie
lower than the UHF CPMC energies, suggesting the CPMC state associated
with the FE trial state is the one most representative of the ground
state.  This state does not show a static striped state. For $U/t=6$ and
$7$ the difference in the FE and UHF CPMC energies is of the order of
$~10^{-2}$ which is likely to be the size of the systematic error.  

Two sets of DMRG results are also presented in Table~1: the lowest
variational bounds obtained (with the largest number of states $m$
used to reach this bound also given) and the extrapolated ground-state
energies $E(P_m=0)$.  For $U/t < 6$ DMRG bounds lie above the FE CPMC
results and the extrapolated DMRG energies agree with the FE CPMC
energies within error estimations.  This demonstrates the accuracy of
both methods in the weak coupling regime.

For $U/t < 6$ neither the CPMC nor DMRG ground states show a static
striped state.  This is consistent with weak coupling renormalization
group calculations which do not show any indication of stripes in
3-chain Hubbard ladders~\cite{lin}.  However, the UHF CPMC results and
some results for metastable DMRG ground states suggest
that low-lying states with stripes exist for relatively small $U$.

Results are significantly different for $U/t \geq 6$.  DMRG
variational bounds are now lower than the CPMC energies, although
differences remain small.  Moreover, DMRG ground states clearly show
striped hole patterns in this regime.  Therefore, these results
suggest that stripes appear in the ground state of the 3-chain Hubbard
model around $U/t=6$.  However, for $U/t \leq 8$ both the FE CPMC
results and DMRG metastable state results show that there are states
without stripes very close to the ground state.  For larger $U/t$
($\geq 12$) DMRG calculations converge very easily to a ground state
with stripes and show little indication of low-lying states without
stripes. CPMC simulations were not attempted for these larger values
of $U$. In Figs.~\ref{fig2} and \ref{fig3}, we show the striped spin
and holes densities for the $U/t=8$ case, along with UHF results for
$U/t=$ and $8$. We note the striking similarity of the CPMC and the
DMRG results with each other and with the $U/t=3$ UHF results, even
though the UHF state has a much higher energy. (The UHF energies for
$U/t=3$ and $8$ are -33.3009 and -19.1907.) From Fig.~2, we identify
two domain walls to be at rung positions 3 and 4 and at 9 and 10. In
each case the spins on either side of the wall are ferromagnetically
aligned but shifted by a phase $\pi$ from what they would be if there
were complete anti-ferromagnetic order. Anti-ferromagnetic order exists between
the walls.  We remark the two-rung domain wall width is similar to
that found by White and Scalapino for the t-J model.

The major import of this work are the qualitative conclusions
that (1) the same characteristic features, indicating a striped
state that were found in the 3-chain t-J model \cite{white97b}, 
are also found in the Hubbard model
at strong to intermediate interaction $U$; and (2) these features
disappear at weaker coupling. These conclusions suggests 
that the striped state may be a new strong-coupling fixed point,
inaccessible to weak-coupling approaches.

Furthermore, the extent to which two quite different numerical
methods agree on a very subtle issue is remarkable.  As used
here, these methods come at the problem with different strengths
and hence mutually calibrate each other.  However, it is crucial
to point out that great care must be used in applying the
methods to avoid coming to erroneous conclusions. In particular,
the CPMC method does not show stripes even at large $U$ when the
FE trial function is used. This underscores the importance of using
trial wavefunctions appropriate to the parameter range of the model,
particularly when the nature of the ground state is
unknown.  

The work of M. G. and J. E. G. was supported by the Department of
Energy.  Some of their calculations used the computers at NERSC.  They
also thank J. Carlson, J. Zaanen, and Shiwei Zhang for helpful comments. J. B.
thanks the Los Alamos National Laboratory for its hospitality while he
performed part of his work. E.J. and S.R.W. acknowledge the support of 
the NSF under Grant No. DMR-9509945 and of 
the Campus Laboratory Collaborations Program of the University of California.

\narrowtext

\begin{table}
\caption{Ground-state energies. For the CPMC results, FE and UHF refer to
  results obtained when either the free-electron or unrestricted Hartree-Fock
  wavefunction was chosen for both the initial and constraining
  wavefunctions. The numbers in parenthesis are the estimated
  statistical error. For the DMRG results the variational energy is
  the one obtained for the maximum number of states kept. This number
  is shown in parenthesis. The extrapolated values are the results
  obtained by the analysis illustrated in Fig.~1. The number in parenthesis
  is an estimate of the error.} 
\begin{tabular}{rdddd}
  &\multicolumn{2}{c}{CPMC}&\multicolumn{2}{c}{DMRG}\\ 
  $U/t$&FE&UHF&Var. bound ($m$)&Extrapolation\\ \tableline
  1&-45.4171 (4) &-45.4161 (4)&-45.3985 (2200)&-45.419 (4)\\ 
  2&-40.581 (2)&-40.560 (2)&-40.5436 (1600)&-40.590 (9)\\ 
  3&-36.622 (1)&-36.566
  (7)&-36.6056 (1600)&-36.627 (4)\\ 4&-33.438 (2)&-33.364 (7)&-33.4259
  (2200)&-33.448 (6)\\ 5&-30.911 (2)&-30.807 (7)&-30.8927 (1600)&-30.915
  (8)\\ 6&-28.873 (4)&-28.807 (6)&-28.8962 (2600)&-28.907 (4)\\ 7&-27.211 
  (5)&-27.133 (6)&-27.2537 (1200)&-27.31 (1)\\ 8&-25.93 (3)&-25.94
  (8)&-25.9502 (1200)&-25.997 (9)\\ 12&-&-&-22.5688 (800)&-22.63 (1)\\ 
  16&-&-&-20.7596 (800)&-20.81 (1)\\ 
\end{tabular}
\label{table1}
\end{table}
 
\begin{figure}
\caption{The DMRG ground-state energy for $U/t=6$ as a function of the
  truncation energy parameterized by $m$, the number of density matrix
  eigenstates.
}
\label{fig1}
\end{figure}

\begin{figure}
\caption{For $U/t=8$, the expectation values of the z-component of the
  rung spin as a function of rung
  position along the middle chain. The solid line is the CPMC result
  using the $U/t=3$ UHF 
  wavefunction as the initial and constraining state. Also shown are
  the UHF results for $U/t=3$ and 8.  The DMRG predicts
  a zero expectation value.
}
\label{fig2}
\end{figure}

\begin{figure}
\caption{For $U/t=8$, the expectation values of the rung hole density
  as a function of rung position.
  The solid line is the CPMC result using the $U/t=3$ UHF wavefunction
  as the initial and constraining state.  The dotted line is the DMRG
  results. Also shown are the UHF results for $U/t=3$ and 8.}
\label{fig3}
\end{figure}

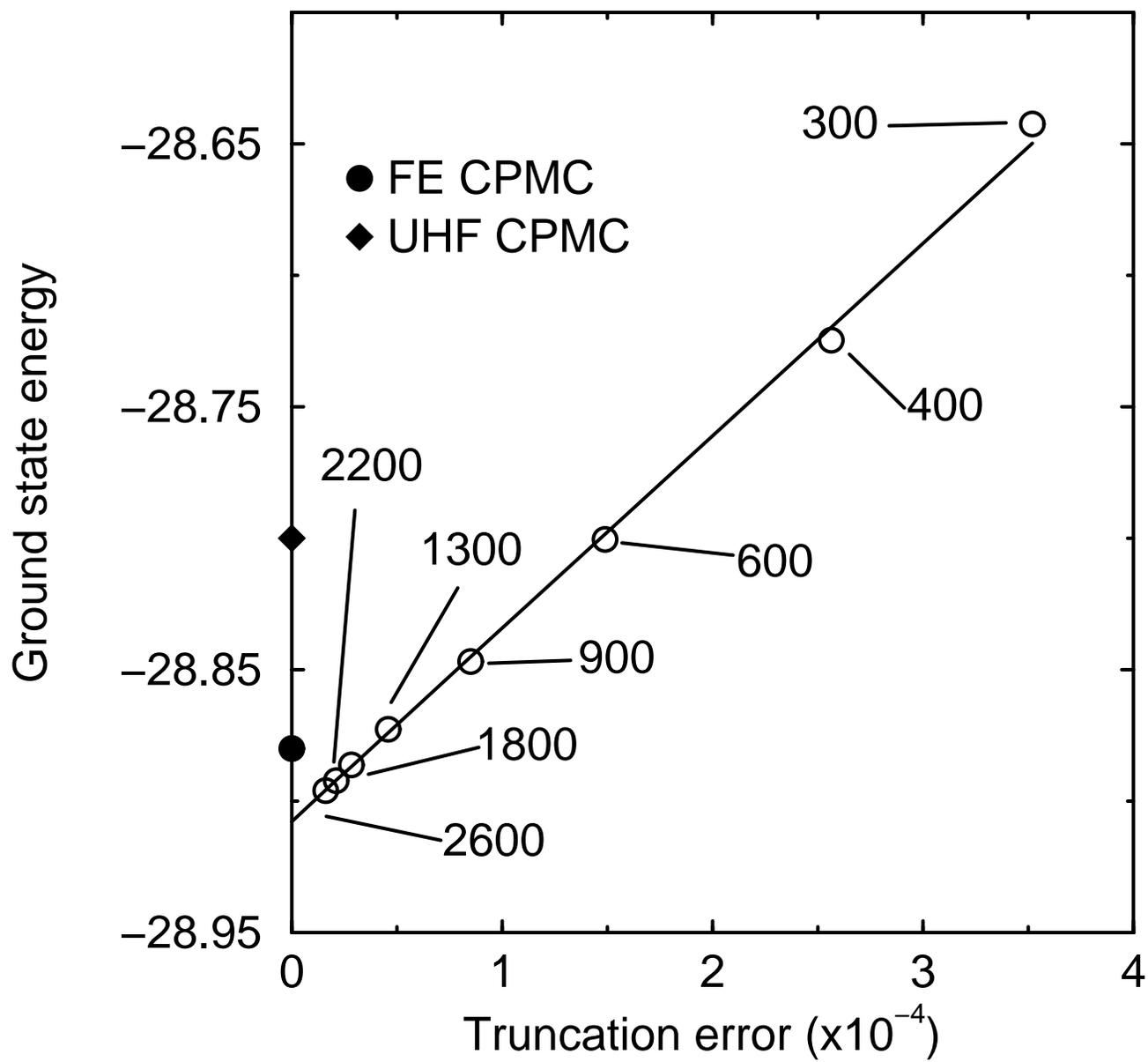
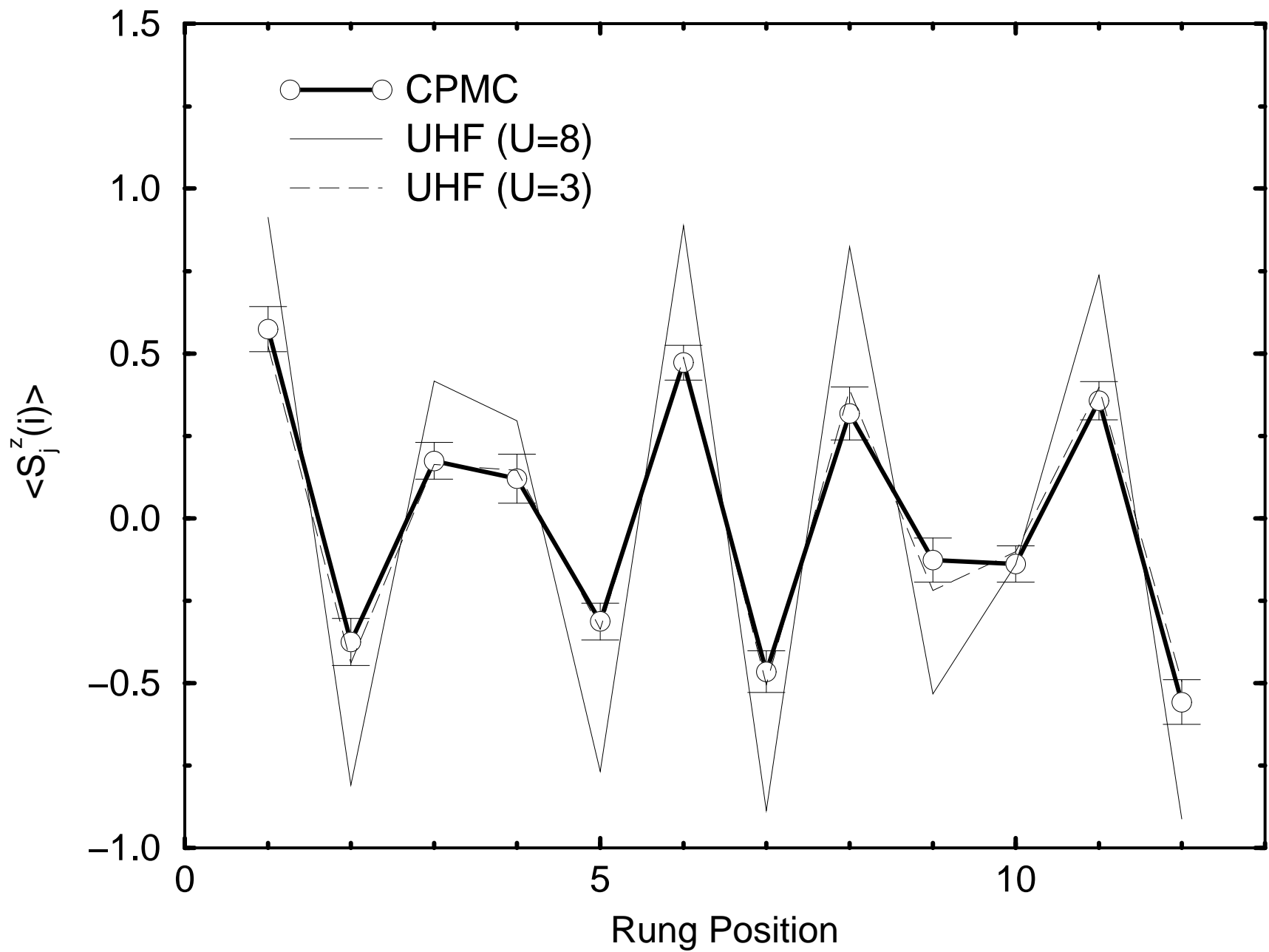
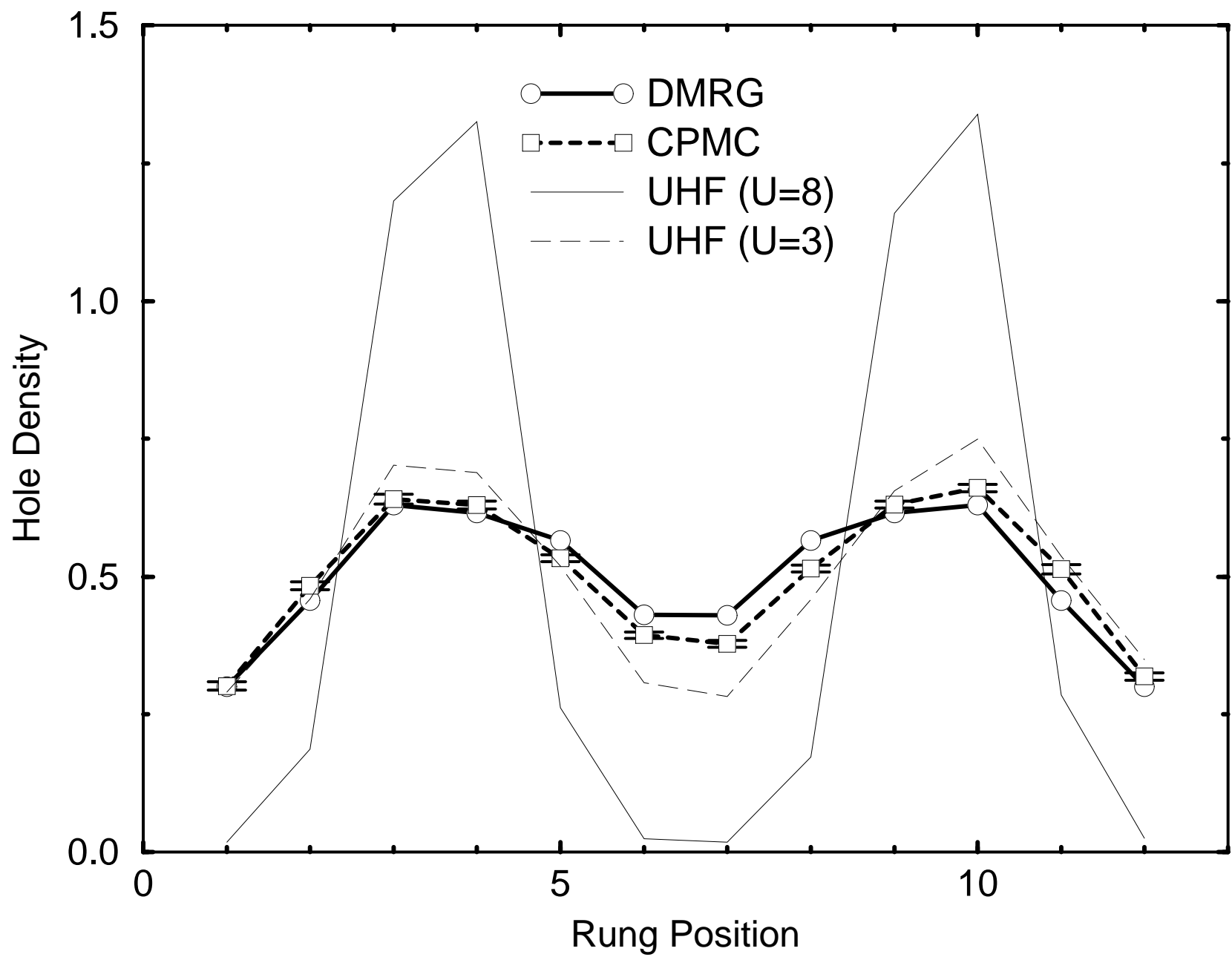
\end{document}